\documentclass[12pt,thmsa]{article}
\usepackage{amsfonts}
\usepackage{amsmath}
\usepackage{amssymb}
\usepackage{sw20lart}

\input{tcilatex}
\begin{document}

\title{Abstract Harmonic Analysis On the non Singular Matrix Lie Group}
\author{Kahar El-Hussein \\
Department of Mathematics, Faculty of Science and Arts at Al Qurayat, \\
Al Jouf University, KSA \ \&\\
Department of Mathematics, Faculty of Science, \\
Al Furat University, Dear El Zore, Syria\\
\textit{\ E-mail: kumath@ju.edu.sa , \ kumath@hotmail.com}}
\maketitle

\begin{abstract}
As well known that it is no way to do the abstract harmonic analysis on the
non connected Lie groups. The goal of this paper is to draw the attention of
Mathematicians to solve this problem. therefore let $\mathbb{R}^{\ast }$ be
the group of nonzero real numbers with multiplication and let $H$ \ be the $%
3-$dimensional Heisenberg group. I denote by $G=H\underset{\rho }{\rtimes }(%
\mathbb{R}^{\ast }$ $)^{3}$ the $6$-dimensional non connected solvable Lie
group, which is isomorphic onto the $6-$ dimensional non singular triangular
matrix Lie group. I define the Fourier-Mellin transform and establish the
Plancherel theorem on the group $G$. Besides I prove the solvability of an
invariant differential operator on the group $G_{+}=H\underset{\rho }{%
\rtimes }(\mathbb{R}_{+}^{\ast }$ $)^{3},$ which is the identity component
of the group $G$. Finally, I give a classification of all left ideals of $%
L^{1}(G_{+})$
\end{abstract}



\section{\protect\bigskip Background}

\textbf{1.1. }The classical Fourier transform is one of the most widely used
mathematical tools in engineering. However, few engineers know that
extensions Fourier analysis on noncommutative Lie groups holds great
potential for solving problems in robotics, image analysis, mechanics.
Engineering applications of noncommutative harmonic analysis brings this
powerful tool to the engineering world. In mathematics, Abstract harmonic
analysis is a generalization of non commutative harmonic analysis in which
results from Fourier analysis are extended to topological groups which are
not commutative. Abstract harmonic analysis, having its roots in the
mid-twentieth century. Its major business is the extension of the theory to
all groups $G$ that are non commutative and locally compact. The results of
its applications are used in the theory of dynamical systems, in the theory
of group representations theory, in the theory of Banach algebra and in many
other mathematical physics theories.

\textbf{1.2. }The best developed branch of abstract harmonic analysis have
been obtained for compact groups by the Peter-Weyl theory for the
representations of compact Lie groups. In the case of non-compact
non-commutative groups the general theory is still far from complete and it
is a difficult task due to the nature of the group representations. There is
still no general theory for approaching the harmonic analysis of an
arbitrary locally compact group.

\textbf{1.3.} In the second half of twenty century, two points of view were
adopted by the community of the mathematics. The first one is the theory of
representations of Lie groups. Unfortunately If the group $G$ is no longer
assumed to be abelian, it is not possible anymore to consider its dual (i.e
the set of all equivalence classes of unitary irreducible representations).
For a long time, people have tried to construct objects in order to
generalize Fourier transform to the non abelian case. However, with the dual
object not being a group, it is not possible to define the Fourier transform
and the inverse Fourier transform between the group $G$ and its dual $G$.
These difficulties of Fourier analysis on noncommutative groups makes the
noncommutative version of the problem very challenging. It was necessary to
find a subgroup or at least a subset of locally compact groups which were
not "pathological", or "wild" as Kirillov calls them $[14]$

The second is the quantum groups (Hopf Algebra), which was introduced by
Vladimir Drinfeld and Michio Jimbo. some little results were obtained by
this theory. Still now neither the theory of quantum groups nor the
representations theory have done to reach this goal.

\textbf{1.4.}\ Since $2006$, and far away of theory of representations of
Lie groups and the of theory of quantum groups, I have opened a new way in
abstract harmonic analysis that no one has opened before.

Therefore, I would like to draw the attention of Scientists in Mathematics
and Physics on the ideas of my way which focus on two approaches:

1- The First one focuses on Fourier transform and Partial differential
equations with variable coefficients on Lie groups. By this way, I have
solved the Lewy and Mizohata operators. I believe that is will be the
business of the expertise in the theory of partial differential equations
with variable coefficients and their applications.

2- By the second way, we can solve the most major problems in Fourier
analysis on many Lie groups. In this paper, I will introduce the abstract
harmonic analysis on the group $G=H\underset{\rho }{\rtimes }(\mathbb{R}%
^{\ast }$ $)^{3}$, which is not only non commutative locally compact group
but is not connected. So I open other new way for Mathematicians and
physicists in the theory of Fourier analysis on a non connected Lie groups

\section{\protect\bigskip \textbf{Introduction} and Results}

\textbf{2.1. }In my book $[10],$ I have proved the set $(\mathbb{R}%
_{-}^{\ast }$ $)^{3}=\{(x,y,z)\in (\mathbb{R}^{\ast })^{3};$ $x$ $\lneqq
0,y\lneqq 0,z\lneqq 0\}$ is group and isomorphic onto the group $(\mathbb{R}%
_{+}^{\ast }$ $)^{3}=\{(x,y,z)\in (\mathbb{R}^{\ast })^{3};$ $x$ $\gneqq
0,y\gneqq 0,z\gneqq 0\}$. I consider the non connected solvable Lie group,
which is consisted of all matrices%
\begin{equation}
G=\left( \left[ 
\begin{array}{ccc}
a_{1}\text{\ } & n_{1} & n_{3} \\ 
0 & a_{1}\text{\ } & n_{2} \\ 
0 & 0 & a_{1}\text{\ }%
\end{array}%
\right] \right) ,(a_{1},a_{2},a_{3}\text{\ })\in (\mathbb{R}^{\ast })^{3}
\end{equation}

\bigskip The $3$-dimensional Heisenberg group $H$, which consists of all
matrices%
\begin{equation}
H=\left( \left[ 
\begin{array}{ccc}
1\text{\ } & n_{1} & n_{3} \\ 
0 & 1\text{\ } & n_{2} \\ 
0 & 0 & 1\text{\ }%
\end{array}%
\right] \right) ,(a_{1},a_{2},a_{3}\text{\ })\in (\mathbb{R}^{\ast })^{3}
\end{equation}%
is normal sub-group of $G.$ The multiplication of two element $%
n=(n_{3},n_{2},n_{1})$ and $m=(m_{3},m_{2},m_{1})$ is given by%
\begin{eqnarray}
n.m &=&(n_{3},n_{2},n_{1})(m_{3},m_{2},m_{1})  \notag \\
&=&(n_{3}+m_{3}+n_{1}m_{2},n_{2}+m_{2},n_{1}+m_{1})
\end{eqnarray}

So the group $G$ can be identified with the group $H\rtimes _{\rho }(\mathbb{%
R}^{\ast })^{3}.$ which is the semi-direct of the group $H$ with group $(%
\mathbb{R}^{\ast })^{3}.$ Since the set $\mathbb{R}_{-}^{\ast }$ is group
isomorphic onto the group $\mathbb{R}_{+}^{\ast }$ see $[10],$ then the
group $\mathbb{R}^{\ast }$ is a two copies of\ the group $\mathbb{R}%
_{+}^{\ast }.$ And so the group $(\mathbb{R}^{\ast })^{3}$ becomes two
copies of the group $(\mathbb{R}_{+}^{\ast })^{3}.$ It is enough to restrict
my study on the group $G_{+}$ of all matrices%
\begin{equation}
G_{+}=\left( \left[ 
\begin{array}{ccc}
a_{1}\text{\ } & n_{1} & n_{3} \\ 
0 & a_{1}\text{\ } & n_{2} \\ 
0 & 0 & a_{1}\text{\ }%
\end{array}%
\right] \right) ,(a_{1},a_{2},a_{3}\text{\ })\in (\mathbb{R}_{+}^{\ast })^{3}
\end{equation}

The group $G_{+}=H\rtimes _{\rho }(\mathbb{R}_{+}^{\star })^{3}$ is the
semi-direct product of $H$ with $(\mathbb{R}_{+}^{\star })^{3},$ where $\rho 
$ is the group homomorphism $\rho :(\mathbb{R}_{+}^{\star
})^{3}\longrightarrow Aut(H)$ defined by: 
\begin{eqnarray}
&&\rho (a)(n_{3},n_{2},n_{1})  \notag \\
&=&\rho
(a_{1},a_{2},a_{2})(n_{3},n_{2},n_{1})=(a_{1}a_{3}^{-1}n_{3},a_{2}a_{3}^{-1}n_{2},a_{1}a_{2}^{-1}n_{1})
\end{eqnarray}

\begin{eqnarray}
&&\rho (a^{-1})(n_{3},n_{2},n_{1})  \notag \\
&=&\rho
(a_{1}^{-1},a_{2}{}^{-1},a_{3}^{-1})(n_{3},n_{2},n_{1})=(a_{1}^{-1}a_{3}n_{3},a_{2}^{-1}a_{3}n_{2},a_{1}^{-1}a_{2}n_{1})
\end{eqnarray}%
for any $a=(a_{1},a_{2},a_{3})\in (\mathbb{R}_{+}^{\star })^{3}$ and $%
(n_{3},n_{2},n_{1})\in H$, where $Aut(H)$ is the group of all automorphisms
of $H.$ The multiplication of two elements $%
X=(n_{3},n_{2},n_{1},a_{1},a_{2},a_{3})$ and $%
Y=(m_{3},m_{2},m_{1},b_{1},b_{2},b_{3})$ in $G_{+}$ is given by 
\begin{eqnarray}
&&X\cdot Y  \notag \\
&=&(n_{3},n_{2},n_{1},a_{1},a_{2},a_{3})(m_{3},m_{2},m_{1},b_{1},b_{2},b_{3})
\notag \\
&=&((n_{3},n_{2},n_{1}).\rho
(a_{1},a_{2},a_{3})(m_{3},m_{2},m_{1}),(a_{1},a_{2},a_{3})(b_{1},b_{2},b_{3}))
\notag \\
&=&((n_{3},n_{2},n_{1}).\rho
(a_{1},a_{2},a_{3})(m_{3},m_{2},m_{1}),(a_{1}b_{1},a_{2}b_{2},a_{3}b_{3})) 
\notag \\
&=&((n_{3},n_{2},n_{1})(a_{1}a_{3}^{-1}m_{3},a_{2}a_{3}^{-1}m_{2},a_{1}a_{2}^{-1}m_{1}),(a_{1}b_{1},a_{2}b_{2},a_{3}b_{3}))
\\
&=&((n_{3}+a_{1}a_{3}^{-1}m_{3}+n_{1}a_{2}a_{3}^{-1}m_{2},n_{2}+a_{2}a_{3}^{-1}m_{2},n_{1}+a_{1}a_{2}^{-1}m_{1}),(a_{1}b_{1},a_{2}b_{2},a_{3}b_{3}))
\notag
\end{eqnarray}

The inverse of an element $X=(n_{3},n_{2},n_{1},a_{1},a_{2},a_{3})$ in $%
G_{+} $ is 
\begin{eqnarray}
X^{-1} &=&(n_{3},n_{2},n_{1},a_{1},a_{2},a_{3})^{-1}  \notag \\
&=&(\rho
((a_{1},a_{2},a_{3})^{-1})((n_{3},n_{2},n_{1}))^{-1},(a_{1},a_{2},a_{3})^{-1})
\notag \\
&=&(\rho
((a_{1},a_{2},a_{3})^{-1})((-n_{3}+n_{1}n_{2},-n_{2},-n_{1})),(a_{1},a_{2},a_{3})^{-1})
\notag \\
&=&((a_{1}^{-1}a_{3}(-n_{3}+n_{1}n_{2}),-a_{2}^{-1}a_{3}n_{2},-a_{1}^{-1}a_{2}n_{1}),(a_{1}^{-1},a_{2}^{-1},a_{3}^{-1}))
\end{eqnarray}

We denote by $C^{\infty }(G),$ $\mathcal{D}(G),$ $\mathcal{D}^{\prime }(G),$ 
$\mathcal{E}^{\prime }(G)$ respectively the space of $C^{\infty }$%
-functions, $C^{\infty }$-functions with compact support, distributions and
distributions with compact support.\ We denote by $L^{1}(G)$ the Banach
algebra that consists of all complex valued functions on the group $G,$
which are integrable with respect to the Haar measure of $G$ and
multiplication is defined by convolution on $G.$ \ 

\textbf{2.2. }Let $\mathcal{U}\;$be the complexified universal enveloping
algebra of the real Lie algebra $\underline{g_{+}}$\ of $G_{+}$; which is
canonically isomorphic to the algebra of all distributions on $G_{+}$
supported by the identity element $(0,0,0,1,1,1)$ of $G_{+}$. For any $u\in 
\mathcal{U}$ one can define a differential operator $P_{u}$ on $G_{+}$ as
follows:%
\begin{equation}
P_{u}f(X)=u\ast f(X)=\int\limits_{G}f(Y^{-1}X)u(Y)dY
\end{equation}%
for any $f\in C^{\infty }(G_{+}),$ where $%
Y=(m_{3},m_{2},m_{1},b_{1},b_{2},b_{3}),$ $dY$ $=dm_{3}dm_{2}dm_{1}\frac{%
db_{1}}{b_{1}}\frac{db_{2}}{b_{2}}\frac{db_{3}}{b_{3}}$ is the right Haar
measure on $G_{+}$, $X=(n_{3},n_{2},n_{1},a_{1},a_{2},a_{3})$ and $\ast $ 
\hspace{0.05in}denotes the convolution product on $G_{+}.$ The mapping $%
u\rightarrow P_{u}$ is an algebra isomorphism of $\mathcal{U}$ onto the
algebra of all right invariant differential operators on $G_{+}$.

\textbf{2.3.} Let $K=$ $H\mathbb{\times }(\mathbb{R}_{+}^{\star })^{3}$ be
the group of the direct product of $H$ and $(\mathbb{R}_{+}^{\star })^{3}$.
We denote also by $\mathcal{U}$ the complexified enveloping algebra of the
real Lie algebra $\underline{k}$ of $K.$ For every $u\in \mathcal{U}$, we
can associate a differential operator $Q_{u}$ on $K$ as follows%
\begin{equation}
Q_{u}f(X)=u\star f(X)=\int\limits_{K}f(Y^{-1}X)u(Y)dY
\end{equation}%
for any $f\in C^{\infty }(K),$ $X\in K,Y\in K.$ where $\star $ signify the
convolution product on the commutative group $K$ and$.$ The mapping $%
u\mapsto Q_{u}$ is an algebra isomorphism of $\mathcal{U}$ onto the algebra
of all invariant differential operators on $K.$ For more details see$[7,13]$%
\ \ 

\section{\protect\bigskip Fourier Transform and Plancherel Theorem on $G_{+}$%
}

\bigskip \textbf{3.1.} Let $L=H\times (\mathbb{R}_{+}^{\star })^{3}\times (%
\mathbb{R}_{+}^{\star })^{3}$ be the group with multiplication 
\begin{eqnarray}
X\cdot Y &=&(n,x,a)(m,y.b)  \notag \\
&=&(n.\rho (a)m,xy,ab)
\end{eqnarray}%
for all $X=(n,x,a)\in L$ and $Y=(m,y,b)\in L,$ where $n=(n_{3},n_{2},n_{1}),$
and $m=(m_{3},m_{2},m_{1})$ The inverse of an element $X=(n,x,a)$ in $L$ is
given by: 
\begin{eqnarray}
X^{-1} &=&(n,x,a)^{-1}  \notag \\
&=&(\rho (a^{-1})n^{-1},x^{-1},a^{-1})
\end{eqnarray}%
where $n^{-1}=(-n_{3}+n_{1}n_{2},-n_{2},-n_{1}),\rho (a^{-1})n^{-1}=\rho
((a_{1},a_{2},a_{3})^{-1})((n_{3},n_{2},n_{1}))^{-1}=(a_{1}^{-1}a_{3}(-n_{3}+n_{1}n_{2}),-a_{2}^{-1}a_{3}n_{2},-a_{1}^{-1}a_{2}n_{1}), 
$ $x^{-1}=(x_{1}^{-1},x_{2}^{-1},x_{3}^{-1}),$ and $%
a^{-1}=(a_{1}^{-1},a_{2}^{-1},a_{3}^{-1}).$ In this case, we can identify $%
G_{+}$ with the closed subgroup $\underbrace{H\times \{1\}\rtimes _{\rho }(%
\mathbb{R}_{+}^{\star })^{3}}$ of $L$ \ \ and $K$ with $H\times (\mathbb{R}%
_{+}^{\star })^{3}\times \{1\}$ of $L.$ From now on, I denote by $an$
instead of $\rho (a)n$

\textbf{Definition 3.1.}\textit{\ For every }$\phi \in C^{\infty }(G_{+})$, 
\textit{one can define a function} $\widetilde{\phi }\in C^{\infty }(L)$ 
\textit{as follows:} 
\begin{equation}
\widetilde{\phi }(n,a,x)=\phi (an,ax)
\end{equation}%
\textit{for all }$(n,a,x)\in L.$

\ \textbf{Remark 3.1.}\textit{\ The function} $\widetilde{\phi }$ \textit{is
invariant in the following sense} 
\begin{equation}
\widetilde{\phi }(bn,ab^{-1},xb)=\widetilde{\phi }(n,a,x)
\end{equation}%
for any $(n,a,x)\in L$ and $b\in (\mathbb{R}_{+}^{\star })^{3}.$ So every
function $\psi (n,a)$ on $G_{+}$ extends uniquely as an invariant function $%
\widetilde{\psi }(n,a,x)$ on $L$

\textbf{Definition 3.2.} \textit{For every} $F\in L^{1}(L)$ \textit{one can
define two convolutions product on the group }$L$ \textit{as}: 
\begin{align*}
& \text{ }g\ast F(n,a,x) \\
& =\int\limits_{G_{+}}F\left[ (m,b)^{-1}(n,a,x)\right] g(m,b)dm\frac{db}{b}
\\
& =\int\limits_{G_{+}}F\left[ ((b^{-1}m^{-1}),b^{-1})(n,a,x)\right] g(m,b)dm%
\frac{db}{b} \\
& =\int\limits_{G_{+}}F\left[ (b^{-1}(m^{-1}n),a,xb^{-1})\right] g(m,b)dm%
\frac{db}{b}
\end{align*}%
\textit{and }%
\begin{align*}
& g\star F(n,a,x_{1}) \\
& =\int\limits_{K}F\left[ ((m^{-1},b^{-1})(n,a,x)\right] g(m,b)dm\frac{db}{b}
\\
& =\int\limits_{K}F\left[ (m^{-1}n,b^{-1}a,x)\right] g(m,b)dm\frac{db}{b}
\end{align*}%
\textit{\ for any }$F\in L^{1}(L),$\textit{where} $a=(a_{1},a_{2},a_{3})\in (%
\mathbb{R}_{+}^{\star })^{3},x=(x_{1},x_{2},x_{3})\in (\mathbb{R}_{+}^{\star
})^{3},b=(b_{1},b_{2},b_{3})\in (\mathbb{R}_{+}^{\star })^{3},$ $dm\frac{db}{%
b}=dm_{3}dm_{2}dm_{1}\frac{db_{1}}{b_{1}}\frac{db_{2}}{b_{2}}\frac{db_{3}}{%
b_{3}}$ \textit{is the right Haar measure on} $G_{+},$ $\ast $ \textit{is
the convolution product on} $G_{+}$ \textit{and} $\star $ \textit{is the} 
\textit{convolution product on} $K.$

It results immediately if $F$ is invariant in sense $(14),$ I get the
following equality%
\begin{equation}
g\ast F(n,a,x)=u\text{ }\star F(n,a,x)
\end{equation}

As in $[10]$, we will define the Fourier-Mellin transform on $G_{+}$.
Therefor let $\mathcal{S}(G_{+})$ be the Schwartz space of $G_{+}$ which%
\hspace{0.05in}can be considered as the Schwartz space of $\mathcal{S}%
(H\times $\ $(\mathbb{R}_{+}^{\star })^{3}),$ and let $\mathcal{S}^{\prime
}(G_{+})$ be the space of all tempered distributions on $G_{+}.$

\bigskip \textbf{Definition 3.3.} \textit{If} $f\in \mathcal{S}(G_{+})$, 
\textit{we define the Fourier transform of its invariant }$\widetilde{f}$ 
\textit{as follows}%
\begin{eqnarray}
&&\mathcal{F}_{H}\mathcal{F}\widetilde{f}(\xi ,\lambda ,1)  \notag \\
&=&\int\limits_{H}\int\limits_{(\mathbb{R}_{+}^{\star })^{3}}\int\limits_{(%
\mathbb{R}_{+}^{\star })^{3}}\int\limits_{\mathbb{R}^{3}}\widetilde{f}(n,a,b)%
\text{ }e^{-\text{ }i\text{ }\xi \text{ }n}\text{ }a^{-i\text{ }\lambda
}b^{-i\mu }dn\frac{da}{a}\text{ }\frac{db}{b}d\mu  \notag \\
&=&\int\limits_{\mathbb{R}^{3}}\mathcal{F}_{H}\mathcal{F}\widetilde{f}(\xi
,\lambda ,\mu )\text{ }d\mu =\mathcal{F}_{H}\mathcal{F}\widetilde{f}(\xi
,\lambda ,1)\text{ }
\end{eqnarray}%
\textit{where }$\mathcal{F}_{H}$ \textit{is the the Fourier transform on the}
$3-$\textit{dimensional Heisenberg group, }$a=(a_{1},a_{2},a_{3}),$ $%
b=(b_{1},b_{2},b_{3}),$ $\frac{da}{a}=\frac{da_{1}}{a_{1}}\frac{da_{2}}{a_{2}%
}\frac{da_{3}}{a_{3}},\frac{db}{b}=\frac{db_{1}}{b_{1}}\frac{db_{2}}{_{b2}}%
\frac{db_{3}}{_{b3}},$ $\lambda =(\lambda _{1},\lambda _{2},\lambda
_{3}),1=(1,1,1),\mu =(\mu _{1},\mu _{2},\mu _{3})$ \textit{and }$d\mu =d\mu
_{1}d\mu _{2}d\mu _{3}$

\textbf{Proposition 3.1 }\textit{\ For every} $g\in \mathcal{S}(G),$ \textit{%
and} $f\in \mathcal{S}(G),$ \textit{we have}

\begin{equation}
\int\limits_{\mathbb{R}^{3}}\mathcal{F}_{H}\mathcal{F}(g\ast \widetilde{f}%
)(\xi ,\lambda ,\mu )d\mu =\mathcal{F}_{H}\mathcal{F}(\widetilde{f})(\xi
,\lambda ,1)\text{ }\mathcal{F}_{H}\mathcal{F}(g)(\xi ,\lambda )
\end{equation}%
\textit{for any }$\xi =(\xi _{1},\xi _{2},\xi _{3})\in H,$ \textit{and} $%
\lambda =(\lambda _{1},\lambda _{2},\lambda _{3})$ $\in \mathbb{R}^{3}$

\bigskip \textit{Proof}: First, we have%
\begin{eqnarray}
&&\ g\ast \widetilde{f}(m,b,x)  \notag \\
&=&\int\limits_{G_{+}}\widetilde{f}((n,a)^{-1}(m,b,x))g(n,a)dn\frac{da}{a} 
\notag \\
&=&\int\limits_{G_{+}}\widetilde{f}(a^{-1}(mn^{-1}),b,xa^{-1})g(n,a)dn\frac{%
da}{a}  \notag \\
&=&\int\limits_{K}\widetilde{f}(mn^{-1},ba^{-1},x)g(n,a)dn\frac{da}{a}%
=g\star \widetilde{f}(m,b,x)\ \ \ 
\end{eqnarray}%
\ \ \ \ \ \ \ \ \ \ \ \ \ \ \ \ \ \ \ \ \ \ \ \ \ \ \ \ \ \ \ 

\ Secondly:%
\begin{eqnarray}
&&\int\limits_{\mathbb{R}^{3}}\mathcal{F}_{H}\mathcal{F}(\overset{\vee }{u}%
\ast \widetilde{f})(\xi ,\lambda ,\mu )d\mu  \notag \\
&=&\int\limits_{\mathbb{R}^{3}}\mathcal{F}_{H}\mathcal{F}(\overset{\vee }{u}%
\star \widetilde{f})(\xi ,\lambda ,\mu )d\mu =\mathcal{F}_{H}\mathcal{F}(%
\widetilde{f})(\xi ,\lambda ,1)\mathcal{F}(\overset{\vee }{u})\text{ }(\xi
,\lambda )\ 
\end{eqnarray}

\textbf{Theorem 3.1. (\textit{Plancheral's Theorem). }}\textit{For any }$%
f\in $\textit{\ }$L^{1}(G_{+})\cap $\textit{\ }$L^{2}(G_{+}),$\textit{we get}

\begin{eqnarray}
&&f\text{ }\ast \widetilde{\overset{\vee }{f}}(0,1,1)  \notag \\
&=&\int\limits_{G_{+}}\left\vert f(n.a)\right\vert ^{2}dn\frac{da}{a}%
=\int\limits_{\mathbb{R}^{6}}\left\vert \mathcal{F}f(\xi ,\lambda
)\right\vert ^{2}d\xi d\lambda
\end{eqnarray}%
\textit{where} $\xi =(\xi _{1},\xi _{2},\xi _{3}),(\lambda _{1},\lambda
_{2},\lambda _{3}),d\xi =d\xi _{1}d\xi _{2}d\xi _{3},d\lambda =d\lambda
_{1}d\lambda _{2}d\lambda _{3},$ \textit{and }$\widetilde{\overset{\vee }{f}}%
(n,a,b)=\widetilde{\overset{\vee }{f}}(an,ab)=\overline{f((an,ab)^{-1})}$

\textit{Proof:}\textbf{\ }First, we have 
\begin{eqnarray}
&&f\text{ }\ast \widetilde{\overset{\vee }{f}}(0,1,1)  \notag \\
&=&\int\limits_{G_{+}}\widetilde{\overset{\vee }{f}}\left[
((a^{-1}n^{-1}),a^{-1})(0,1,1)\right] f(n,a)dn\frac{da}{a}  \notag \\
&=&\int\limits_{G_{+}}\widetilde{\overset{\vee }{f}}%
((a^{-1}n^{-1}),1,a^{-1})f(n,a)dn\frac{da}{a}  \notag \\
&=&\int\limits_{G_{+}}\overset{\vee }{f}((a^{-1}n^{-1}),a^{-1})f(n,a)dn\frac{%
da}{a}  \notag \\
&=&\int\limits_{G_{+}}\overline{f(n,a)}f(n,a)dn\frac{da}{a}%
=\int\limits_{G_{+}}\left\vert f(n,a)\right\vert ^{2}dn\frac{da}{a}
\end{eqnarray}

\bigskip Secondly by $(14)$, we get 
\begin{eqnarray}
&&f\text{ }\ast \widetilde{\overset{\vee }{f}}(0,1,1)  \notag \\
&=&\int\limits_{H}\ \int\limits_{\mathbb{R}^{6}}\mathcal{F}_{H}\mathcal{F}%
(f\ast \widetilde{\overset{\vee }{f}})(\xi ,\lambda ,\mu )d\xi d\lambda d\mu
=\int\limits_{\mathbb{R}^{5}}\mathcal{F}(f\star \widetilde{\overset{\vee }{f}%
})(\xi ,\lambda ,\mu )d\xi d\lambda d\mu  \notag \\
&=&\int\limits_{H}\ \int\limits_{\mathbb{R}^{3}}\mathcal{F}(\widetilde{%
\overset{\vee }{f}})(\xi ,\lambda ,1)\mathcal{F}(f)\text{ }(\xi ,\lambda
)d\xi d\lambda =\int\limits_{H}\ \int\limits_{\mathbb{R}^{3}}\overline{%
\mathcal{F}_{H}\mathcal{F}(f)\text{ }}(\xi ,\lambda )\mathcal{F}(f)\text{ }%
(\xi ,\lambda )d\xi d\lambda  \notag \\
&=&\int\limits_{H}\ \int\limits_{\mathbb{R}^{3}}\left\vert \mathcal{F}_{H}%
\mathcal{F}(f)\text{ }(\xi ,\lambda )\right\vert ^{2}d\xi d\lambda
=\int\limits_{G_{+}}\left\vert f(n,a)\right\vert ^{2}dn\frac{da}{a}
\end{eqnarray}

So the Plancheral theorem on\textit{\ }$G_{+}.$

\textbf{Corollary 3.1. }\textit{For any }$f\in L^{2}(G=H\rtimes _{\rho }(%
\mathbb{R}^{\ast })^{3}),$ I get\textbf{\ }%
\begin{eqnarray}
&&\int\limits_{G}\left\vert f(n,(a_{1},a_{2},a_{3}))\right\vert ^{2}dn\frac{%
da_{1}}{a_{1}}\frac{da_{2}}{a_{2}}\frac{da_{3}}{a_{3}}  \notag \\
&=&2\int\limits_{\mathbb{R}^{3}}\left\vert \mathcal{F}_{H}\mathcal{F}(f)%
\text{ }(\xi ,(\lambda _{1},\lambda _{2},\lambda _{3}))\right\vert ^{2}d\xi
d\lambda _{1}d\lambda _{2}d\lambda _{3}
\end{eqnarray}%
\textit{where }$n=(n_{3},n_{2},n_{1}),$ $\xi =(\xi _{3},\xi _{2},\xi _{1}).$ 
\textit{and} $d\xi =d\xi _{3}d\xi _{2}d\xi _{1}$

\textit{Proof: }Since the group $G=H\rtimes _{\rho }(\mathbb{R}^{\ast })^{3}$
is a two copies of the group $H\rtimes _{\rho }(\mathbb{R}_{+}^{\ast })^{3}.$
That means $H\rtimes _{\rho }(\mathbb{R}^{\ast })^{3}=H\rtimes _{\rho }(%
\mathbb{R}_{+}^{\ast })^{3}\cup H\rtimes _{\rho }(\mathbb{R}_{+}^{\ast
})^{2},$ so I obtain%
\begin{eqnarray}
&&f\text{ }\star \widetilde{\overset{\vee }{f}}(0,1,1)=\int\limits_{G}\left%
\vert f(n,(a_{1},a_{2},a_{3}))\right\vert ^{2}dn\frac{da_{1}}{a_{1}}\frac{%
da_{2}}{a_{2}}\frac{da_{3}}{a_{3}}  \notag \\
&=&2\int\limits_{G_{+}}\left\vert f(n,(a_{1},a_{2},a_{3}))\right\vert ^{2}dn%
\frac{da_{1}}{a_{1}}\frac{da_{2}}{a_{2}}\frac{da_{3}}{a_{3}}  \notag \\
&=&2\int\limits_{\mathbb{R}^{6}}\left\vert \mathcal{F}_{H}\mathcal{F}(f)%
\text{ }(\xi ,(\lambda _{1},\lambda _{2},\lambda _{3}))\right\vert ^{2}d\xi
d\lambda _{1}d\lambda _{2}d\lambda _{3}
\end{eqnarray}

Hence the proof of the corollary

\section{Left Ideals of the Group Algebra $L^{1}(G_{+})$}

\bigskip First, I will prove the solvability of any invariant differential
operator on the connected solvable group $G_{+}.$ Therefore denote by $%
\widetilde{C^{\infty }(G_{+})}$ $(resp.$ $\widetilde{C^{\infty }(K)\text{ }}$%
) the image of $C^{\infty }(G_{+}$ $)$ $(resp.C^{\infty }(K)$\ then we have%
\begin{equation*}
\ \widetilde{C^{\infty }(G_{+})}|_{G_{+}}=C^{\infty }(G_{+})
\end{equation*}%
\begin{equation}
\ \widetilde{C^{\infty }(K)}|_{K}=C^{\infty }(K)
\end{equation}

\bigskip \textbf{Definition 4.1. }\textit{Let }$\chi $ \textit{be the mapping%
} $:\widetilde{C^{\infty }(K)}|_{K\text{ }}$ $\longrightarrow \widetilde{%
C^{\infty }(G_{+})}|_{G_{+}\text{ }}$\textit{\ defined by}%
\begin{equation}
\widetilde{f}|_{K}\ (n,a,1)\rightarrow \widetilde{f}|_{G_{+}}(n,1,a)
\end{equation}%
\textit{is topological isomorphism and its inverse is nothing but }$\chi
^{-1}$\textit{defined by}

\begin{equation}
\widetilde{f}|_{G_{+}}\ (n,1,a)\rightarrow \widetilde{f}|_{K}(n,a,1)
\end{equation}

My main result is

\textbf{Theorem 4.1.} \textit{If }$P_{u}$\ \textit{any} \textit{invariant
differential operator on }$G_{+}$\textit{\ associated to the distribution }$%
u\in \mathcal{U}$\textit{, then, we have }

\begin{equation}
P_{u}\text{ }C^{\infty }(G_{+})=C^{\infty }(G_{+})
\end{equation}

\textit{Proof:} Let $Q_{u}$\ be the invariant differential operator on $K$
associated to $u$ , then by the theory of my book \ for the invariant
differential operators on the Heisenberg group $H$ and the theory of partial
differential operators with constant coefficients $[23]$, we get 
\begin{equation}
Q_{u}\ \ \widetilde{C^{\infty }(K)}|_{K}=\widetilde{C^{\infty }(K)}%
|_{K}=C^{\infty }(K)
\end{equation}

That means for any $\psi (n,a)\in C^{\infty }(K),$ there exist a function $%
\varphi (n,a,x)\in \widetilde{C^{\infty }(K)},$ such that%
\begin{equation}
Q_{u}\varphi (n,a,1)=u\star \varphi (n,a,1)=\psi (n,a)
\end{equation}

The function $\psi (n,a)$ can be\ transformed as an invariant function $\psi
\in \widetilde{C^{\infty }(K)}$ as follows

\begin{equation}
\psi (n,a)=\widetilde{\psi }((a^{-1}n),a,1)
\end{equation}

In other side, we have%
\begin{eqnarray}
&&\chi Q_{u}\ \ \varphi (n,a,1)  \notag \\
&=&Q_{u}\ \ \varphi (n,1,a)=u\star \varphi (n,1,a)  \notag \\
&=&u\ast \varphi (n,1,a)=P_{u}\text{ }\varphi (n,1,a)  \notag \\
&=&\chi \widetilde{\psi }(a^{-1}n,a,1)=\widetilde{\psi }(a^{-1}n,1,a)  \notag
\\
&=&\psi (n,a)
\end{eqnarray}%
where 
\begin{equation}
u\star \varphi (n,1,a)=\left\{ \int\limits_{K}\varphi \left[
m^{-1}n,b^{-1},a)\right] u(m,b)dm\frac{db}{b},\text{ }\varphi \in \text{ }%
\widetilde{C^{\infty }(K)}\right\}
\end{equation}%
and%
\begin{equation}
u\ast \varphi (n,1,a)=\left\{ \int\limits_{G_{+}}\varphi \left[
(b^{-1}(m^{-1}n),1,ab^{-1})\right] g(m,b)dm\frac{db}{b},\text{ }\varphi \in 
\text{ }\widetilde{C^{\infty }(G_{+})}\right\}
\end{equation}

So the proof of the solvability of any right invariant differential operator
on $G_{+}.$

If $I$ is a subspace of $L^{1}(G_{+}),$ we denote $\widetilde{I}$ its image
by the mapping $\thicksim $ let $J=$ $\widetilde{I}\ |_{K}.$ My main result
is:

\textbf{Theorem 4.2.} \textit{Let} $I$ \textit{be a subspace of} $%
L^{1}(G_{+}),$ \textit{then the following conditions are equivalents}.

$(i)$ $J=\widetilde{I}\ |_{K}$ \textit{is a left ideal in the Banach algebra}
$L^{1}(K).$

$(ii)$ $I$ \textit{is a left ideal in the Banach algebra} $L^{1}(G_{+}).$

\bigskip \textit{Proof:} $(i)$ implies $(ii)$\ Let $I$ be a subspace of the
space $L^{1}(L)$ such that $J=\widetilde{I}|_{K}$ \ is a left ideal in $%
L^{1}(K),$ then we have: 
\begin{equation}
u\star \widetilde{I}\ |_{K}(n,a,1)\subseteq \widetilde{I}\ |_{K}(n,a,1)
\end{equation}%
for any $u\in L^{1}(K)$ and $(n,a)\in K$, where%
\begin{equation}
u\star \widetilde{I}\ |_{K}(n,a,1)=\left\{ \int\limits_{K}\widetilde{f}%
|_{K}\ \left[ m^{-1}n,ab^{-1},1)\right] u(m,b)dm\frac{db}{b},\text{ }f\in 
\text{ }I\right\}
\end{equation}

It shows that%
\begin{equation}
u\star \widetilde{f}\ |_{K}(n,a,1)\in \widetilde{I}\ |_{K}(n,a,1)
\end{equation}%
for any $\widetilde{f}\in \widetilde{I}.$ Then we get%
\begin{eqnarray}
&&\chi (u\star \widetilde{f}|_{K})(n,a,1)  \notag \\
&=&u\text{ }\ast \widetilde{f}(n,1,a)\in \widetilde{I}\ |_{G_{+}}(n,1,a)=I
\end{eqnarray}%
$(ii)$ implies $(i),$ if $I$ is an ideal in $L^{1}(G_{+}),$ then we get 
\begin{eqnarray}
&&u\ast \widetilde{I}\ |_{G_{+}}(n,1,a)  \notag \\
&=&u\ast I\ (n,a)\subseteq \widetilde{I}\ |_{G_{+}}(n,1,a)=I\ (n,a)
\end{eqnarray}%
where%
\begin{equation}
u\ast \widetilde{I}\ |_{G_{+}}(n,1,a)=\left\{ \int\limits_{G_{+}}\widetilde{f%
}|_{G_{+}}\ \left[ (b^{-1})(m^{-1}n),1,ab^{-1}\right] u(m,b)dm\frac{db}{b},%
\text{ }f\in \text{ }I\right\}
\end{equation}

So, we obtain%
\begin{eqnarray}
&&\chi ^{-1}(u\ast \widetilde{f}\ |_{G_{+}})(n,1,a)\in \chi ^{-1}(\widetilde{%
I}\ |_{G_{+}})(n,1,a)  \notag \\
&=&\widetilde{I}\ |_{K}(n,a,1)
\end{eqnarray}%
and%
\begin{equation}
\chi ^{-1}(u\ast \widetilde{f}\ |_{G_{+}})(n,1,a)=u\star \widetilde{f}%
|_{K}(n,a,1)\in \widetilde{I}\ |_{K}(n,a,1)
\end{equation}

\textbf{Corollary 4.1}. \textit{Let }$I$ \textit{be a subspace of the space} 
$L^{1}(G_{+})$ \textit{and} $\widetilde{I}$ \textit{its image by the mapping}
$\thicksim $ \textit{such that} $J=\widetilde{I}|_{K}$ \textit{is an} 
\textit{ideal in} $L^{1}(K),$ \textit{then the following conditions are
verified}.

$(1)$ $J$ \textit{is a closed left ideal in the algebra} $L^{1}(K)$ \textit{%
if and only if} $I$ \textit{is a closed left ideal in the algebra }$%
L^{1}(G_{+}).$

$(2)J$ \textit{is a prime left ideal in the algebra} $L^{1}(K)$ \textit{if
and only if} $I$ \textit{is a prime left ideal in the algebra }$L^{1}(G_{+})$

$(3)J$ \textit{is a maximal left ideal in the algebra} $L^{1}(K)$ \textit{if
and only if} $I$ \textit{is a maximal left ideal in the algebra }$%
L^{1}(G_{+})$

$(4)$ $J$ \textit{is a left dense ideal in the algebra} $L^{1}(K)$\textit{\
if and only if }$I$\textit{\ is a dense left ideal in the algebra} $%
L^{1}(G_{+}).$

The proof of this corollary results immediately from theorem \textbf{3.2.}

The Heisenberg group $H$ is the semi-direct product of the two vector Lie
group $\mathbb{R}^{2}\rtimes _{\sigma }\mathbb{R}$, where $\sigma :\mathbb{%
R\rightarrow }Aut(\mathbb{R}^{2})$ is the homomorphism group defined. I
extend the group $K=H\mathbb{\times }(\mathbb{R}_{+}^{\ast })^{3}$ by
considering the new group $S$ $=$ $\mathbb{R}^{2}\times \mathbb{R\times
R\times }(\mathbb{R}_{+}^{\ast })^{3}$ with the following law%
\begin{eqnarray}
&&X\cdot Y  \notag \\
&=&(n_{3},n_{2},n_{1},n_{4},a_{1},a_{2},a_{3})(m_{3},m_{2},m_{1},m_{4},b_{1},b_{2},b_{3})
\notag \\
&=&((n_{3}+\sigma
(n_{4})(m_{3},m_{2}),n_{2}+m_{2},n_{1}+m_{1},n_{4}+m_{4}),(a_{1}b_{1},a_{2}b_{2},a_{3}b_{3}))
\notag \\
&=&((n_{3}+m_{3}+n_{4}m_{2},n_{2}+m_{2},n_{1}+m_{1},n_{4}+m_{4}),(a_{1}b_{1},a_{2}b_{2},a_{3}b_{3}))
\end{eqnarray}

Denote by $B=$ $\mathbb{R}^{2}\times \mathbb{R\times }(\mathbb{R}_{+}^{\ast
})^{3}$ the commutative Lie group of the direct product of three Lie groups $%
\mathbb{R}^{2},\mathbb{R}$, and $\ (\mathbb{R}_{+}^{\ast })^{3}.$ In this
case the group $K=H\mathbb{\times }(\mathbb{R}_{+}^{\ast })^{3}$ can be
identified with the sub-group $\mathbb{R}^{2}\times \mathbb{\{}0\mathbb{%
\}\times R\times }(\mathbb{R}_{+}^{\ast })^{3}$ and the group $B=\mathbb{R}%
^{2}\times \mathbb{R\times }(\mathbb{R}_{+}^{\ast })^{3}$ can be identified
with the sub-group $\mathbb{R}^{2}\times \mathbb{R\times \{}0\mathbb{%
\}\times }(\mathbb{R}_{+}^{\ast })^{3}$

\textbf{Definition 4.2. }\textit{Any function }$\psi \in C^{\infty }(K)$ 
\textit{can be extended to a unique function }$\Upsilon \psi $ \textit{%
belongs to }$C^{\infty }(S),$ as follows%
\begin{eqnarray}
&&\Upsilon \psi ((n_{3},n_{2},n_{1},n_{4}),(x_{1},x_{2},x_{3}))  \notag \\
&=&\psi ((\sigma (n_{1})(n_{3},n_{2}),n_{1}+n_{4}),(x_{1},x_{2},x_{3})) 
\notag \\
&=&\psi ((n_{1}(n_{3},n_{2}),n_{1}+n_{4}),(x_{1},x_{2},x_{3}))  \notag \\
&=&\psi ((n_{3}+n_{1}n_{2},n_{2},n_{1}+n_{4}),(x_{1},x_{2},x_{3}))
\end{eqnarray}%
\textit{for any }$(n_{3},n_{2},n_{1},n_{4})\in H\times \mathbb{R}%
,x=(x_{1},x_{2},x_{3})\in (\mathbb{R}_{+}^{\ast
})^{3},n_{1}(n_{3},n_{2})=(n_{3}+n_{1}n_{2},n_{2})=\sigma
(n_{1})(n_{3},n_{2}).$ Note here the function $\Upsilon \psi $ is invariant
in the following sense%
\begin{eqnarray}
&&\Upsilon \psi ((n_{3},n_{2},n_{1},n_{4}),(x_{1},x_{2},x_{3}))  \notag \\
&=&\Upsilon \psi ((\sigma
(m)(n_{3},n_{2}),m^{-1}n_{1},mn_{4}),(x_{1},x_{2},x_{3}))
\end{eqnarray}

If $I$ is a subspace of $L^{1}(K),$ we denote $\Upsilon I$ its image by the
mapping $\Upsilon $. Let $J=$ $\Upsilon I\ |_{B}.$

My main results are:

\textbf{Theorem 4.3.} \textit{Let} $I$ \textit{be a subspace of} $L^{1}(K),$ 
\textit{then the following conditions are equivalents}.

$(i)$ $J=\Upsilon I\ |_{B}$ \textit{is an ideal in the commutative Banach
algebra} $L^{1}(B).$

$(ii)$ $I$ \textit{is a left ideal in the Banach algebra} $L^{1}(K).$

For the proof of this theorem, I refer to my book $[10,$ $ChapI,theorem$ $%
3.1.$ $]$\ \ \ 

\textbf{Corollary 4.2}. \textit{Let }$I$ \textit{be a subspace of the space} 
$L^{1}(K)$ \textit{and} $\Upsilon I$ \textit{its image by the mapping} $%
\Upsilon $ \textit{such that} $J=\Upsilon I|_{B}$ \textit{is an} \textit{%
ideal in} $L^{1}(K),$ \textit{then the following conditions are verified}.

$(1)$ $J$ \textit{is an ideal in the commutative algebra }$L^{1}(B)$ \textit{%
if and only if} $I$ \textit{is a closed left ideal in the algebra }$L^{1}(K)$
\textit{if and only if} $I$ \textit{is a closed left ideal in the algebra }$%
L^{1}(G_{+}).$

$(2)J$ \textit{is a prime ideal in the commutative algebra} $L^{1}(B)$ 
\textit{if and only if} $I$ \textit{is a prime left ideal in the algebra }$%
L^{1}(K)$ \textit{if and only if} $I$ \textit{is a prime left ideal in the
algebra }$L^{1}(G_{+})$

$(3)J$ \textit{is a maximal ideal in the commutative algebra }$L^{1}(B)$ 
\textit{if and only if} $I$ \textit{is a maximal left ideal in the algebra }$%
L^{1}(K)$ \textit{if and only if} $I$ \textit{is a left maximal ideal in the
algebra }$L^{1}(G_{+})$

$(4)$ $J$ \textit{is a dense ideal in the commutative algebra }$L^{1}(B)$%
\textit{\ if and only if }$I$\textit{\ is a dense left ideal in the algebra} 
$L^{1}(K)$ \textit{if and only if }$I$\textit{\ is a left dense ideal in the
algebra} $L^{1}(G_{+})$

The proof of this corollary results as a consequence from theorems \textbf{%
4.3}. and \textbf{4.2}.

\section{\protect\bigskip Discussion}

\bigskip \textbf{5.1. }In this paper I have proved that any invariant
differential operator has the form%
\begin{equation}
P=\sum\limits_{\alpha ,\beta }a_{\alpha \beta }X^{\alpha }Y^{\beta }
\end{equation}%
on the Lie group $G_{+}$, where $X^{\alpha }=(X_{1}^{\alpha
_{1}},X_{2}^{\alpha _{2}},X_{3}^{\alpha _{3}}),Y^{\beta }=(Y_{1}^{\beta
_{1}},Y_{2}^{\beta _{2}},Y_{3}^{\beta _{3}}),$ $\alpha _{i}\in 
\mathbb{N}
,\beta _{i}\in 
\mathbb{N}
$ $(1\leq i\leq 3),$ and $X=(X_{1},X_{2},X_{3}),Y=(Y_{1},Y_{2},Y_{3})$ are
the invariant vectors field on $G_{+},$ which are the basis of the Lie
algebra \underline{$g$}$_{+}$ of $G_{+}$ and $a_{\alpha ,\beta }\in 
\mathbb{C}
.$ As in theorem 5.1. any non zero invariant partial differential equation
on the $6-$dimensional group $G_{+}$ is solvable.

\textbf{5.2. }If we take the special case the $3-$dimensional Heisenberg
group. Any invariant differential operator has the form6%
\begin{equation}
P=\sum\limits_{\alpha ,\beta ,\gamma }a_{\alpha \beta \gamma }X^{\alpha
}Y^{\beta }Z^{\gamma }
\end{equation}%
on the Lie group $H=\mathbb{R}^{2}\times _{\rho }\mathbb{R}$, where $\alpha
\in 
\mathbb{N}
,\beta \in 
\mathbb{N}
$ $,\gamma \in 
\mathbb{N}
,$ and $X,Y,$ and $Z$ , are the invariant vectors on $H,$ which are the
basis of the Lie algebra \underline{$h$} of $H$ and $a_{\alpha \beta \gamma
}\in 
\mathbb{C}
.$ Hence $P$ is solvable. In particular the Lewy operator $X+iY$ is solvable$%
.$ For Mathematicians, there are many non solved problems in Fourier
analysis on this group.

\section{Conclusion.}

\bigskip \textbf{6.1. }It is well known the solvability of the invariant
differential operators and the ideals of the group algebra of a Lie group
play an important role in mathematical analysis and mathematical physics. In
this paper I have defined the Fourier transform and established the
Plancherel theorem on the non connected group $G.$ A classification of all
left ideals in Banach algebra $L^{1}(G_{+})$ is obtained and the solvability
of any non zero invariant differential operator on the group $G_{+}$

\section{\protect\bigskip Competing Interests:}

\bigskip \textbf{7.1. }All Mathematicians interest in the theory of abstract
harmonic analysis (non commutative harmonic analysis) to solve the major
problems in analysis on non connected and non commutative Lie groups. As the
Plancherel theorem, the solvability of the invariant differential operators
on Lie groups. Also the study of the non commutative group algebra of the
Lie groups and their ideals

\section{Author Contributions}

\textbf{8.1. }Far a way from the representations theory, my contributions
is: I have opened a new way in abstract harmonic analysis in order to do the
Fourier analysis on Lie groups. The first one of my goals is the solvability
and hypoellipticity of the invariant partial differential equations on many
Lie groups as Nilpotent, Semi-simple, Lorentz group, Poincare group, ...
ect. This leads us to study the non commutative group algebra and their
ideals of the Lie groups . My discovery of new groups leads me to my second
goal which is the Fourier analysis on non connected Lie groups. According to
this goal I open other new way, which will be the business of the expertise
in Mathematics and Mathematical Physics

\end{document}